\documentclass{article}
\usepackage{spconf,amsmath,graphicx}

\usepackage{amsmath}
\usepackage{amscd}
\usepackage{amsfonts}
\usepackage{mathtools}
\usepackage{epsfig}
\usepackage{theorem}
\usepackage[margin=0.5in]{geometry}

\def \C {\mathbb C}

\title{Spectral Perturbations of the Line Graph Laplacian}
\name{Ian Rooney, Parker Kuklinski, and David Hague \thanks{This research was funded by ONR grant N0001419WX00754}}
\address{Naval Undersea Warfare Center \\
		1176 Howell St., Newport, RI 02841}
\begin{document}
%
\maketitle
\begin{abstract}
The graph Laplacian is an important tool in Graph Signal Processing (GSP) as its eigenvalue decomposition acts as an analogue to the Fourier transform and is known as the Graph Fourier Transform (GFT).  The line graph has a GFT that is a direct analogue to the Discrete Cosine Transform Type II (DCT-II).  Leveraging Fourier transform properties, one can then define processing operations on this graph structure that is loosely analogous to processing operations encountered in Digital Signal Processing (DSP) theory.  This raises the question of whether well defined DSP-like operations can be derived from the GFT for more complex graph structures.  One potential approach to this problem is to perturb simple graph structures and study the perturbation's impact on the graph Laplacian.  This paper explores this idea by examining the eigenvalue decomposition of the Laplacian of undirected line graphs that undergo a single edge weight perturbation. This single perturbation can perturb either an existing edge weight or create new edge between distant unconnected vertices.  The eigenvalue bounds are expressed in closed form and agree with simulated examples.  The theory can be extended to include multiple perturbations such that the GFT can be defined for a more general graph structure. 
\end{abstract}
\section{Introduction}
\label{sec:intro}

Graph Signal Processing (GSP) aims to develop data models and processing techniques for data or signals defined on graphs \cite{OrtegaI, OrtegaII}.  One of the most fundamental processing techniques for signals on graphs is the Graph Fourier Transform (GFT).  The GFT is the eigenvalue decomposition of the graph Laplacian matrix $L$, a matrix defined as the difference between the degree matrix $D$ and the adjacency matrix $A$ of a graph such that
\begin{equation}
L=D-A.
\label{eq:graphLaplacian}
\end{equation}The GFT eigenvectors represent in an abstract sense a notion of frequency of the data on a graph with each subsequent eigenvector representing a higher frequency component.  Using Fourier transform properties developed in standard Digital Signal Processing (DSP) theory, the graph Fourier components derived from the GFT can be used to define various DSP-like processing operations on a graph.  

Graphs with a specific structure have well defined GFTs from which DSP-like operations can be well defined.  For example, line and ring graphs have GFTs that reduce to the Type-II Discrete Cosine Transform (DCT-II) and Discrete Fourier Transform (DFT) respectively \cite{LineDCT, RingDFT}.  In fact, the ring graph defines the standard integer shift-based linear time invariant signals and systems commonly encountered in standard DSP theory.  The basic linear time-invariant DSP operations are mathematically well defined on the ring graph structure.  These operations also have a firm physical interpretation of the impact these operations have on the data that resides on the graph.  However, for irregular graphs with more complex structure, the mathematics and the physical interpretation of the GFT become more obscured.  For example, the spacing the of GFT frequencies can become non-uniform and irregular with some GFT frequencies having high multiplicity \cite{OrtegaII}.  Additionally, much of the GFT analysis has been performed for undirected graphs while frequency definitions for directed graphs are less explored \cite{Chung, Bauer} though new results on directed graphs are starting to appear in the literature \cite{sevi2018harmonic}.  

Nevertheless, the GFT can still be a powerful tool for a multitude of graph structures that are based on simpler graph structures.  For example, introducing perturbations by either adding new edges or modifying existing edges to simple line and ring graphs can substantially modify the structure of those simple graphs.  These perturbations will in turn modify the structure of the graph Laplacian.  Therefore if the eigenvalue decomposition for a perturbed version of the graph Laplacian are well defined then the GFT may be well defined also.  A well defined GFT for these general graph structures should allow more rigorously and clearly defined DSP-like operations on said graph structures.  This paper explores this idea by examining the eigenvalue decomposition of the Laplacian of line graphs that undergo a single edge weight perturbation between either existing edge weights or creating new edges between distant vertices.  The eigenvalues are bounded according to a first order asymptotic approximation.  These bounds are expressed in closed form and agree with simulation.  Moreover, this theory can be extended to include additional perturbations to model the GFT of a more general graph structure.

The idea of introducing spectral perturbations to simple graph structures is not new.  Work by Saito et. al \cite{SaitoI, SaitoII} studied the phase transitions of the eigenvalues of the Laplacian of starlike tree graphs (i.e, a line graph that has exactly one vertex of degree higher than 2) by perturbing a single vertex of a line graph.  More recently, work by Poignard et. al \cite{poignard2018spectra} analyzed the spectra of graphs whose existing edges were perturbed while keeping the general graph structure unchanged.  This research differs from the results in \cite{SaitoI, SaitoII, poignard2018spectra} in that the perturbation can be applied to either existing edges or creating edges and finds closed form expressions for the asymptotic expansions of the perturbed eigenvalues.  The rest of this paper is organized as follows : Section \ref{sec:lineGraphPerturb} derives the spectrum of the simple line graph and its perturbed versions and specifically derives a closed form expression for the eigenvalue bounds of the perturbed line graph Laplacian.  Section \ref{sec:examples} provides computational examples verifying the closed form results  in Section \ref{sec:lineGraphPerturb} and demonstrates the behavior of the eigenvectors of a perturbed line graph Laplacian.  Lastly, Section \ref{sec:Conclusion} concludes the paper.

\section{Perturbation of the Line Graph Laplaican}
\label{sec:lineGraphPerturb}

In this section we outline the derivation of the eigenvalues of the first order perturbation of the line graph Laplacian. First we use Jacobi's formula to derive a formula for the first order perturbation of the eigenvalues of a general graph \cite{Kato}. We then calculate the well known eigenvalues of the line graph Laplacian, and then apply the first order perturbation equation to this graph.

\subsection{Derivation of First Order Matrix Perturbation}
\label{subsec:perturbDerive}
Let $A,M\in\C ^{n\times n}$ be $n\times n$ matrices with $A$ invertible and containing only simple eigenvalues. In this section we are interested in a first order approximation of the spectrum of the matrix $A(\epsilon )=A+\epsilon M$ with $\epsilon\rightarrow 0$. Since $A$ has only simple eigenvalues, its characteristic polynomial satisfies $p'(\lambda )\ne 0$ for all eigenvalues $\lambda\in\sigma (A)$. By the inverse function theorem, we can write the following approximation of the eigenvalues $\lambda (\epsilon )$ of $A(\epsilon )$:
\begin{equation}
\lambda (\epsilon )=\lambda (0)+\epsilon\lambda '(0)+O(\epsilon ^2)
\label{eq:perturbedLGeigvals}
\end{equation}

To compute the value $\lambda '(0)$, we take a derivative of characteristic equation with respect to $\epsilon$:
\begin{equation}
\det (\lambda (\epsilon )I-A(\epsilon ))=0
\end{equation}
Recall Jacobi's formula \cite{Magnus} which allows us to differentiate a determinant
\begin{equation}
\frac{d}{dt}\det A(t)=\text{tr}(\text{adj}[A(t)]A'(t))
\end{equation}
where the adjugate matrix $\text{adj}[A]$ is the transpose of the cofactor matrix. Applying Jacobi's formula to the characteristic equation gives the result
\begin{equation}
\text{tr}\left(\text{adj}[\lambda (\epsilon )I-A(\epsilon )](\lambda '(\epsilon )I-A'(\epsilon ))\right) =0.
\end{equation}
Since the trace operator is linear, we can write
\begin{align}
\lambda '(\epsilon )=\frac{\text{tr}\left(\text{adj}[\lambda (\epsilon )I-A(\epsilon )]A'(\epsilon )\right)}{\text{tr}\left(\text{adj}[\lambda (\epsilon )I-A(\epsilon )]\right)}
\end{align}

\subsection{Spectrum of the Line Graph Laplacian}
\label{subsec:lineGraphLaplacianI}

Consider a line graph with vertices $\{ v_1,...,v_n\}$ and edge weights $w(v_i,v_{i+1})=1$ for $i\in\{ 1,...,n-1\}$. Let $L_n$ be the corresponding graph Laplacian and let $f_n(\lambda )$ be the corresponding characteristic polynomial of $L_n$.  The graph Laplacian is the difference between the degree matrix and the adjacency matrix.  We also consider an auxillary matrix $H_n$ with characteristic polynomial $g_n(\lambda )$ and write these matrices as
\begin{align}
L_n &= \begin{bmatrix} 1 & -1 & ~ & ~ & ~ \\ -1 & 2 & -1 & ~ & ~ \\ ~ & -1 & \ddots & \ddots & ~ \\ ~ & ~ & \ddots & 2 & -1 \\ ~ & ~ & ~ & -1 & 1\end{bmatrix},\\ H_n &= \begin{bmatrix} 2 & -1 & ~ & ~ & ~ \\ -1 & 2 & -1 & ~ & ~ \\ ~ & -1 & \ddots & \ddots & ~ \\ ~ & ~ & \ddots & 2 & -1 \\ ~ & ~ & ~ & -1 & 1\end{bmatrix}.
\end{align}
Via a cofactor expansion, we can prove
\begin{align}
f_k(\lambda ) &= (\lambda -1)g_{k-1}(\lambda )-g_{k-2}(\lambda ), \\ g_k(\lambda ) &= (\lambda -2)g_{k-1}(\lambda )-g_{k-2}(\lambda )
\end{align}
Representing this recursion as a matrix multiplication
\begin{equation}
\begin{bmatrix} g_{k+1} \\ g_k\end{bmatrix} =\begin{bmatrix} \lambda -2 & -1 \\ 1 & 0\end{bmatrix}\begin{bmatrix} g_k \\ g_{k-1}\end{bmatrix}
\end{equation}
and using the initial conditions $g_{-1}(\lambda )=-1$ and $g_0(\lambda )=1$, an eigenvalue expansion of this matrix leads to a closed form for $f_n(\lambda )$:
\begin{equation}
f_n(\lambda )=\frac{\lambda F_{n}(\lambda )}{F_1(\lambda )}.
\end{equation}
Here, $F_n(\lambda )=\omega _+(\lambda )^n-\omega _-(\lambda )^n$ and $\omega _\pm (\lambda )$ is defined as
\begin{equation}
\omega _\pm (\lambda )=\frac{1}{2}\left[\lambda -2\pm\sqrt{(\lambda -2)^2-4}\right].
\end{equation} If we make the substitution $\lambda =2\cos\theta +2$, then $\omega _\pm (\lambda )=e^{i\theta}$
such that
\begin{equation}
f_n(\theta )=(2\cos\theta +2)\frac{\sin n\theta}{\sin\theta}.
\end{equation}
We can index solutions of this equation as $\theta _k=\frac{\pi k}{n}$ for $k\in\{ 1,...,n\}$ such that the eigenvalues become
\begin{equation}
\lambda _k=2\cos\left(\frac{\pi k}{n}\right) +2,
\label{eq:LGeigvals}
\end{equation}
which corresponds to the DCT-II \cite{LineDCT, Strang}. 

\subsection{Spectrum of Perturbed Line Graph Adjacency Matrix}
\label{subsec:perturbedLineGraphLaplacian}

Now suppose that we perturb the weight between $v_{m_1}$ and $v_{m_2}$ by $\epsilon$ and that we wish to compute $\lambda _k'(0)$ with respect to this perturbation. This perturbation on the graph Laplacian can be written as $A(\epsilon )=L_n+\epsilon M$ where $M$ satisfies $(M)_{m_1m_1}=(M)_{m_2m_2}=1$, $(M)_{m_1m_2}=(M)_{m_2m_1}=-1$, and is zero elsewhere.  We first compute the denominator of (1), which using the properties of the adjoint matrix can be written as
\begin{equation}
\text{tr}\left(\text{adj}[\lambda _k(0)I-A(0)]\right) =\sum _{i=1}^n [\lambda _k(0)I-A(0)]_{ii}
\end{equation}
where $[A]_{ij}$ is the determinant of the ${ij}$ cofactor matrix of $A$. By inspection of the matrix $A$, we see that this cofactor sum satisfies:
\begin{equation}
\sum _i [\lambda _kI-A]_{ii}=\sum _{i=1}^ng_{i-1}(\lambda _k)g_{n-i}(\lambda _k).
\end{equation}
Using the trigonometric identities
\begin{align}
\sum _{k=1}^n\sin k\theta &=\frac{\sin (n\theta /2)\sin ((n+1)\theta /2)}{\sin (\theta /2)}, \\ \sum _{k=1}^n\cos k\theta &= \frac{\sin (n\theta /2)\cos ((n+1)\theta /2)}{\sin (\theta /2)}
\end{align}
and the angle-sum identities, we can simplify (16) as follows:

\begin{align*}
  \sum _i [\lambda _kI-L_n]_{ii}  = & \frac{1}{\sin ^2\theta _k}\sum _{i=1}^n\left(\sin i\theta +\sin (i-1)\theta\right)\times \\ {}& \left(\sin (n-i+1)\theta +\sin (n-i)\theta\right),\\
\begin{split}
    {}&= -(-1)^k\frac{1+\cos\theta _k}{\sin ^2\theta _k}\\
      &\times\left[n-\sin\theta _k\left(\sum _{i=1}^n\sin 2i\theta _k\right) \right. \\ & \left. -\cos\theta _k\left(\sum _{i=1}^n\cos (2i\theta _k)\right)\right],
\end{split}\\
    {}&= -\frac{(-1)^kn(1+\cos\theta _k)}{\sin ^2\theta _k} 
\end{align*}

\begin{equation}
\sum _i [\lambda _kI-L_n]_{ii}  =-\frac{2(-1)^kn\cos ^2(\theta _k/2)}{\sin ^2\theta _k}.
\end{equation}
To proceed, consider $M$ as a sum of four single entry matrices, i.e. $M=M_{m_1m_1}+M_{m_2m_2}-M_{m_1m_2}-M_{m_2m_1}$. In this way, we can substitute $M$ for $A'(0)$ in equation (1) and simplify the trace calculation. We have $\text{tr }(XM_{ij})=(X)_{ji}$ since $(XM_{ij})_{jj}=(X)_{ji}$ and the matrix vanishes elsewhere on the diagonal. If $X=\text{adj }Y$, then $(X)_{ji}=(-1)^{i+j}[Y]_{ij}$. It is possible to show that for the graph Laplacian, this cofactor satisfies:
\begin{equation}
[\lambda I-L_n] _{m_1m_2}=g_{m_1-1}(\lambda )g_{n-m_2}(\lambda )
\end{equation}
Using several trigonometric identities gives the final expression:
\begin{multline}
\lambda '(0)=\frac{2}{n}\left[ (-1)^{m_1}\sin\left(\left( m_1-\frac{1}{2}\right)\frac{\pi k}{n}\right) - \right. \\ \left. (-1)^{m_2}\sin\left(\left( m_2-\frac{1}{2}\right)\frac{\pi k}{n}\right)\right]^2.
\label{eq:lambdaprime}
\end{multline}

\section{Some Illustrative Examples}
\label{sec:examples}

This section details a few key examples that illustrate and validate the eigenvalue derivations from the previous sections. Additionally, this section presents numerical simulations pertaining to the eigenvectors of a perturbed line graph Laplacian.

\subsection{Eigenvalues}

\begin{figure}[ht]
\centering
\includegraphics[width=0.5\textwidth]{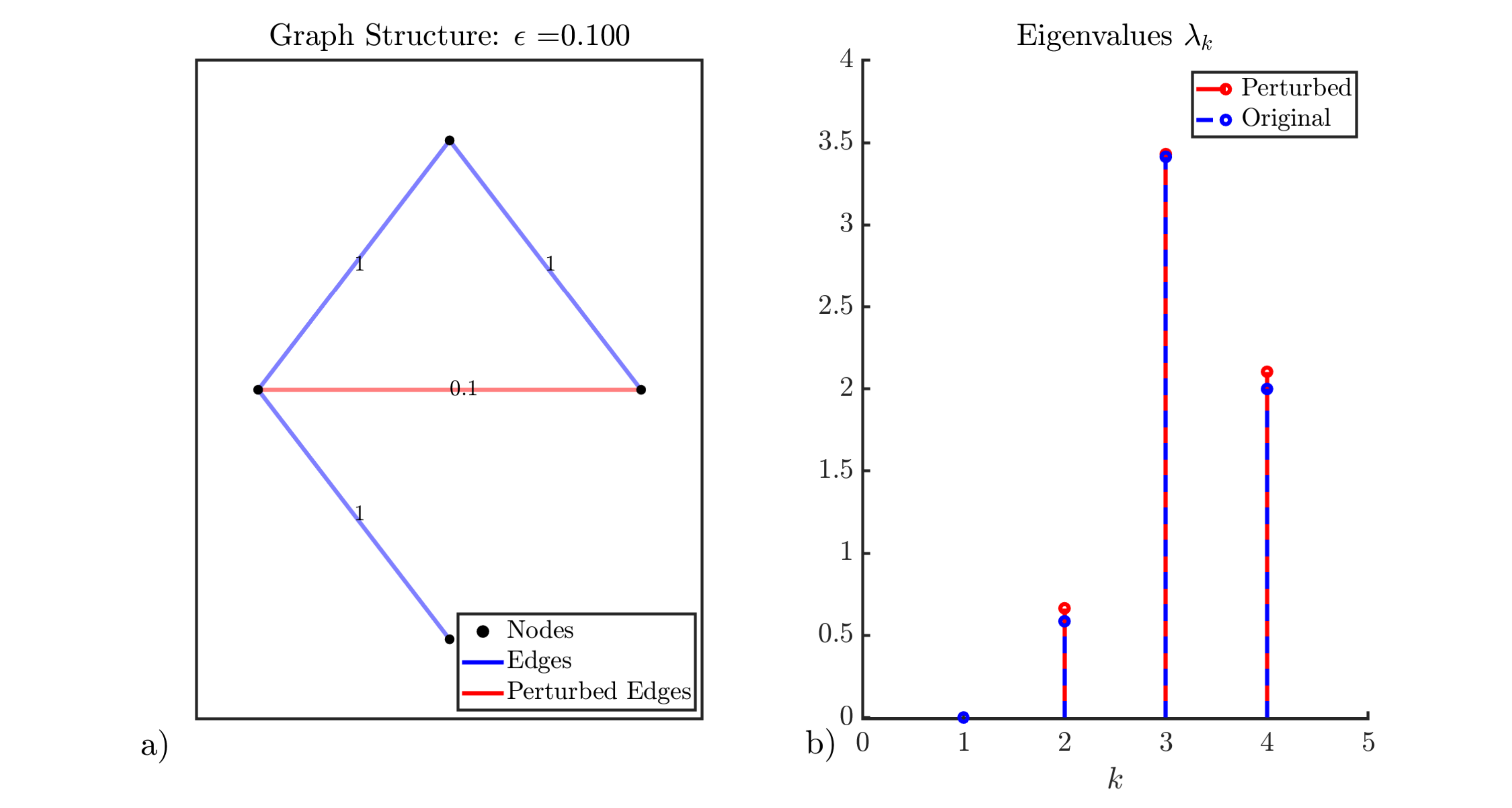}
\caption{a) A four-node line graph (blue) with equal edge weights. This line graph is perturbed by adding a new edge (red) with weight $\epsilon$ between nodes 2 and 4. b) The eigenvalues of the graph Laplacian of the graphs shown in part a. The eigenvalues of the perturbed graph (red) are within an $O(\epsilon^2)$ error of the predicted eigenvalues in \eqref{eq:perturbedLGeigvals}.}
\label{fig:EigValA}
\end{figure}

Consider the simple graph structure shown in blue in Figure~\ref{fig:EigValA}a. It is a 4 node line graph with equal edge weights. The exact eigenvalues of this graph Laplacian $L$ are given by \eqref{eq:LGeigvals} such that $\sigma[L]=\{0,0.586,3.414,2\}$, which are depicted in blue on Figure~\ref{fig:EigValA}b. We then apply a single pertubation, shown in red, with edge weight $\epsilon=0.1$ between nodes $2$ and $4$ of the line graph. The numerically computed eigenvalues are plotted in red on Figure~\ref{fig:EigValA}b. These numerically computed perturbed eigenvalues are in close agreement with the perturbed eigenvalues predicted by \eqref{eq:perturbedLGeigvals},~\eqref{eq:LGeigvals}, and~\eqref{eq:lambdaprime} within an error consistent with an $O(\epsilon^2)$ term.

\begin{figure}[ht]
\centering
\includegraphics[width=0.5\textwidth]{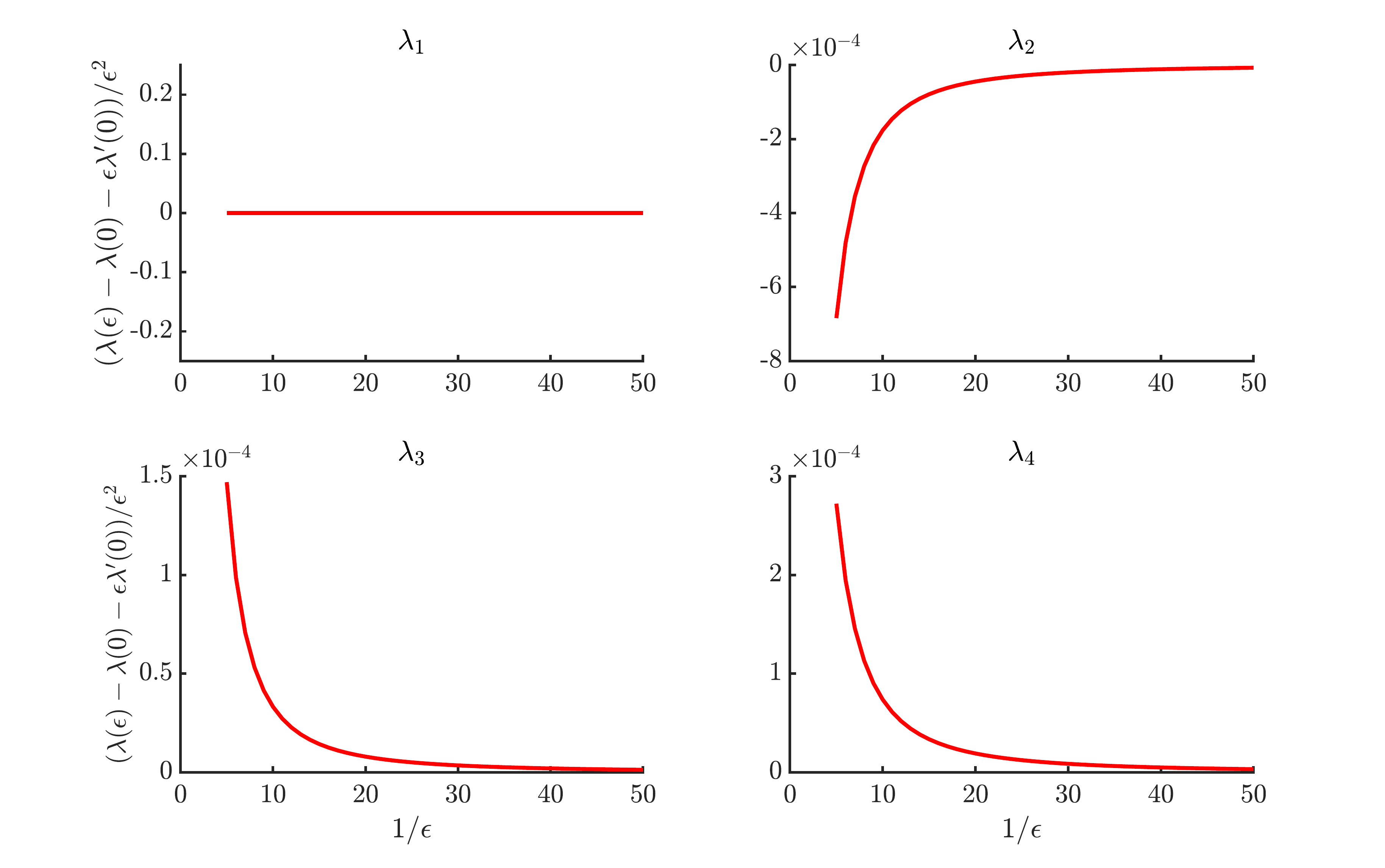}
\caption{The numerically computed error term given by \eqref{eq:eigvalError} for each eigenvalue of the perturbed graph structure in Figure~\ref{fig:EigValA}a as a function of $\epsilon$. The functions each converge to $0$ in the infinite limit of decreasing $\epsilon$. This validates that \eqref{eq:perturbedLGeigvals} is a accurate approximation of the eigenvalues for this graph stucture.}
\label{fig:EigValB}
\end{figure}

Next, consider the same perturbed graph structure as Figure~\ref{fig:EigValA}a with  $0.02 \leq \epsilon \leq 0.2$. Equation \eqref{eq:perturbedLGeigvals} predicts the perturbed eigenvalues as a function of $\epsilon$, but the prediction has an $O(\epsilon^2)$ term dictating an error term proportional to $\epsilon^2$. By rearranging \eqref{eq:perturbedLGeigvals}, we can express the error $E$ as
\begin{equation}
E=\frac{\lambda (\epsilon )-\left(\lambda (0)+\epsilon\lambda '(0)\right)}{\epsilon ^2}.
\label{eq:eigvalError}
\end{equation}
Ideally, $E$ should converge to a constant in the limit of decreasing $\epsilon$. Figure~\ref{fig:EigValB} depicts the results of \eqref{eq:eigvalError} using the numerically computed actual eigenvalues for each value of $\epsilon$ in red. For each eigenvalue of this graph structure, $E$ appears to converge to $0$ in the limit of decreasing $\epsilon$. This means that the $O(\epsilon^2)$ term in \eqref{eq:perturbedLGeigvals} is negligible for small values of $\epsilon$ and the first order approximation is a good approximation. Furthermore, in the infinite limit as $\epsilon$ approaches zero, the perturbed eignvalues approach the original, unperturbed eigenvalues of the four-node line graph structure.

\subsection{Eigenvectors}

\begin{figure}[ht]
\centering
\includegraphics[width=0.5\textwidth]{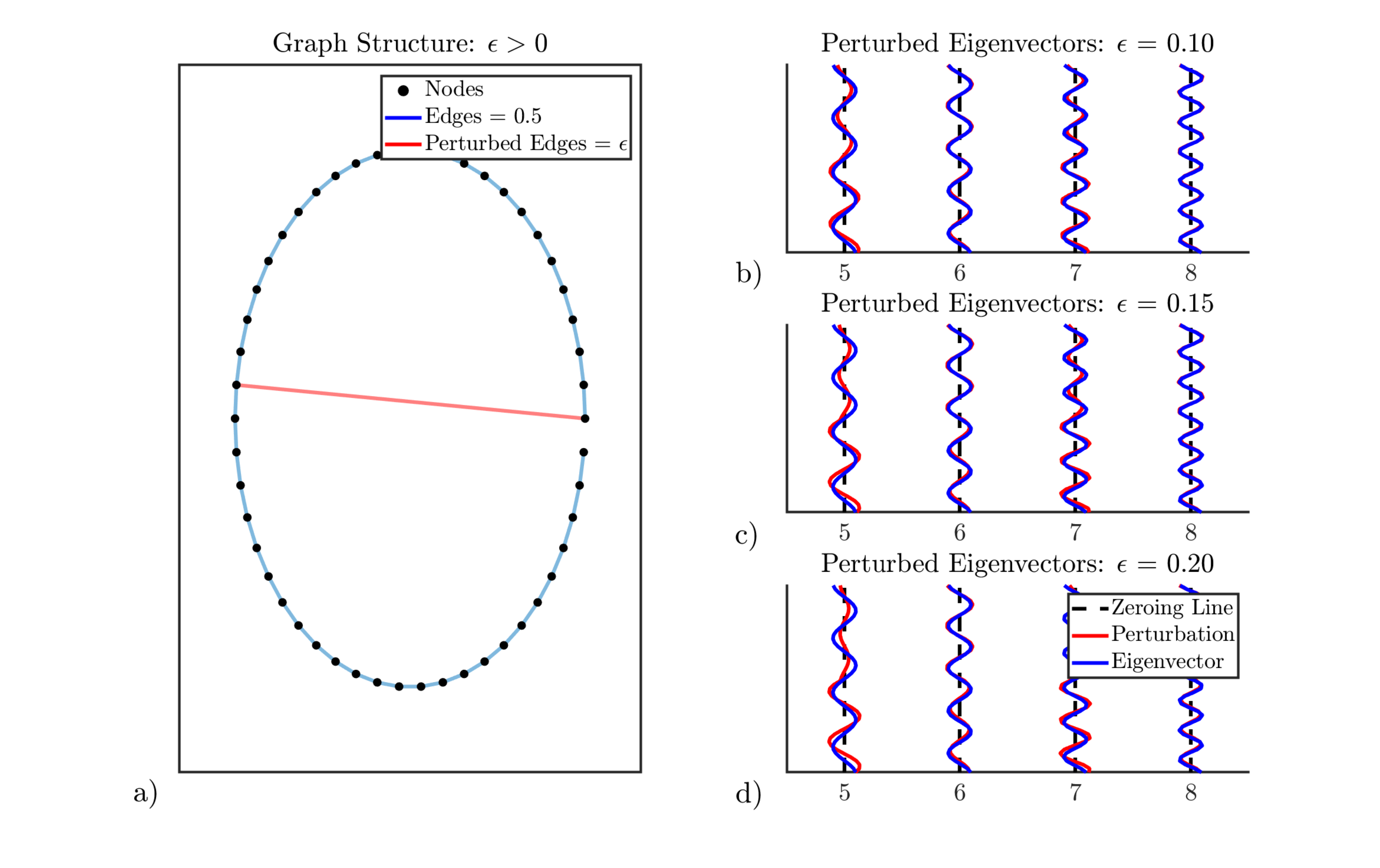}
\caption{a) A fifty-node line graph (blue) with equal edge weights perturbed by adding a new edge (red) with weight $\epsilon$ between nodes 26 and 50. b-d) The fifth through eighth eigenvectors of the graph Laplacian are shown in part a for $\epsilon=[0.1,0.15,0.2]$, respectively. The eigenvectors transition smoothly between the eigenvectors of the unperturbed graph (blue) and the eigenvectors of the perturbed graph (red).
}
\label{fig:animationFrame}
\end{figure}

Although the majority of this paper focuses on eigenvalues, eigenvectors are also an important piece of graph signal processing. Perturbations of eigenvectors are more complicated to compute analytically than the eigenvalues and are best relegated to numerical computation.  Simulations have shown that the magnitude of eigenvectors of a perturbed undirected line graph transition smoothly over $\epsilon$, with some minor caveats. For the purposes of this paper, ``smooth transition'' refers to the property of the eigenvectors' magnitudes displaying no large discontinuities between the $n^{th}$ eigenvector of the line graph with perturbation weight $\epsilon$ and the $n^{th}$ eigenvector of the same perturbed graph with pertubation weight $\epsilon\pm\Delta$, where $\Delta$ is an arbitrarily small number. Two animated examples of this phenomena can be found in this paper's supplemenatal files. A graphical example is depicted by Figure~\ref{fig:animationFrame}.

There are however some caveats to the smooth transition of eigenvectors assumption. First, both the original and perturbed line graph structure must have simple eigenvalues. That is, all eigenvalues must have multiplicity 1. Second, eigenvectors are subject to sign flips. The  $n^{th}$ eigenvector for $\epsilon$ and the $n^{th}$ eigenvector for $\epsilon+\Delta$ may be $\pi$ radians out of phase with each other and thus $\textbf{v}_n(\epsilon)\approx -\textbf{v}_n(\epsilon+\Delta)$. In numerical computation, these sign changes can be easily identified and corrected to maintain a visually smooth transition.

\vspace{-2em}
\section{Conclusion}
\label{sec:Conclusion}

This paper derived a first order asymptotic approximation for the eigenvalues of an undirected line graph Laplacian perturbed by a singular edge with weight $\epsilon$. The analytical expression is validated with numerical results that show the error term approaching a constant value. The eigenvectors of such a perturbed line graph are shown to transition smoothly between arbitrarily small changes in $\epsilon$.  There are several avenues to pursue that build from the results described in this paper.  These are deriving closed-form expressions for the eigen-vectors, introducing multiple perturbations to model the GFT of a more general graph structure, and extending this analysis for the ring graph and directed graphs.  The end goal of these extensions are to develop a stronger intuition for the behavior of DSP-like operations defined on more general graph structures.



\end{document}